\documentclass[12pt]{article}
\usepackage{hyperref}
\usepackage{amsmath}
\usepackage{amssymb}
\usepackage{amsthm}
\usepackage{graphicx}
\usepackage{verbatim}

\newtheorem{lemma}{Lemma}
\newcommand{\fig}[3]{
\begin{figure}
\centerline{
\includegraphics[width=200pt]{figures/#1.pdf}
}
\caption{#3}
\label{#2}
\end{figure}
}
\newcommand{\figsize}[4]{
\begin{figure}
\centerline{
\includegraphics[width=#1]{figures/#2.pdf}
}
\caption{#4}
\label{#3}
\end{figure}
}

\def\del{\partial}

\title{Rotation distance using flows}
\author{Claire Mathieu \and William P. Thurston}
\date{1992
\footnote{
AUTHOR'S NOTE:
Bill Thurston and I composed this extended abstract
in 1992.
The proof is an outline that was meant to be
completed in a subsequent full paper.
This work was set aside,
and after a time no complete copy of this abstract was known to exist.
A copy was recently found, and I am making it available now,
for information purposes.\\
--- Claire Mathieu, 25 August 2024
}
}

\begin{document}

\maketitle

\begin{abstract}
Splay trees are a simple and efficient dynamic data structure,
invented by Sleator and Tarjan.
The basic primitive for transforming a binary tree in this scheme
is a rotation.
Sleator, Tarjan, and Thurston proved that
the maximum rotation distance between trees with $n$ internal nodes
is exactly $2n-6$
for trees with $n$ internal nodes
(where $n$ is larger than some constant).
The proof of the upper bound is easy
but the proof of the lower bound,
remarkably,
uses sophisticated arguments
based on calculating hyperbolic volumes.
We give an elementary proof of the same result.
The main interest of the paper lies in the method,
which is new.
It basically relies on a potential function argument,
similar to many amortized analyses.
However, the potential of a tree is not defined explicitly,
but by constructing an instance of a flow problem
and using the max-flow min-cut theorem.
\end{abstract}

\vspace{1cm}

\section{Introduction} \label{sec1}

\paragraph*{Background.}
Splay trees are a simple and efficient dynamic data structure,
invented by Sleator and Tarjan
[1].
The basic primitive for transforming a binary tree in this scheme
is a \emph{rotation} (see figure \ref{fig1}).
\fig{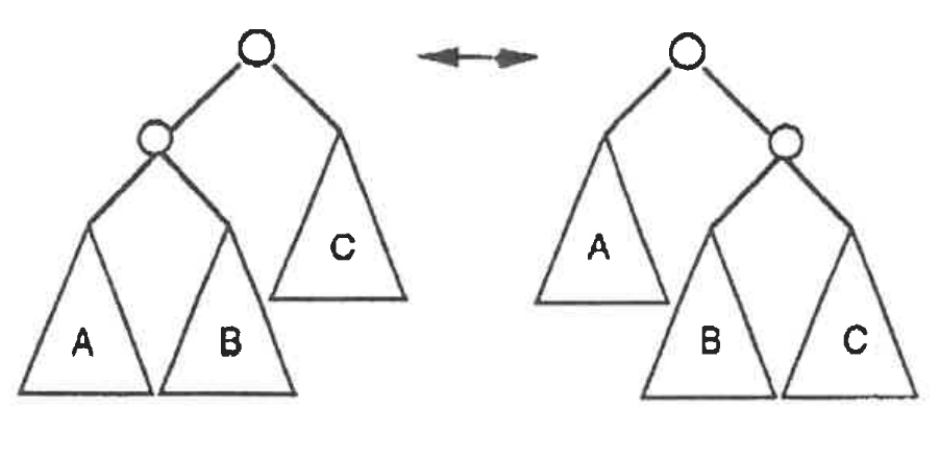}{fig1}{Rotating an edge of a binary tree.}
The way in which splay trees evolve
during a sequence of transformations
is still not well understood,
as witnessed by the resistance of the dynamic optimality conjecture
to all efforts so far.
In an attempt to understand how rotations act on trees,
Sleator, Tarjan and Thurston addressed the question
of computing the rotation distance between binary trees,
i.e. the minimum number of rotations required
to transform a given tree into another given tree
[2].
They proved that the maximum distance is exactly
$2n-6$
for trees with $n$ internal nodes
(where $n$ is larger than some constant).
The proof of the upper bound is easy
but the proof of the lower bound,
remarkably,
uses sophisticated arguments
based on calculating hyperbolic volumes.

\paragraph*{Results.}
In this paper,
we give an elementary proof of the same result,
using the max-flow min-cut theorem
instead of hyperbolic geometry.
One advantage of the proof is that
we exhibit two ``natural trees''
which are at maximum distance apart,
namely two variants of the ``zig-zag'' tree.
(In the original proof,
the two trees are not explicitly defined).

\paragraph*{The proof method.}
The main interest of the paper lies in the method,
which is new.
It basically relies on a potential function argument,
similar to many amortized analyses.
However, the potential of a tree is not defined explicitly
(as is usually the case in an amortized analysis),
but by constructing an instance of a flow problem
and using the max-flow min-cut theorem.
It is also interesting to see an example of a problem
where potential functions can be used
for proving lower bounds and not only upper bounds.

\paragraph*{Comparing the two methods.}
Our proof has the obvious advantage of being purely combinatorial,
so that it is essentially self-contained.

One drawback is that it involves some tedious case-by-case analysis.
We must stress that this tedious work is unnecessary
for proving a lower bound of $2n-O(1)$:
if we do not insist in proving
that the maximum distance is exactly $2n-6$,
but are happy with the less precise lower bound $2n-O(1)$,
our proof technique works much more smoothly,
and most of the case-by-case analyses disappear.
In contrast,
the proof of [2],
which works by successive refinements of the lower bound,
is relatively easy for a lower bound of $2n-O(\sqrt{n})$,
but would require almost as much work
to prove $2n-O(1)$ as $2n-6$.

For other potential applications,
we think that our method is more powerful.
Instead of relying on the hyperbolic metric,
in which solving the problem consists in
positioning $n$ points in space adequately
(with $3n$ degrees of freedom),
our approach is akin to constructing our own metric,
geared to the problem which we wish to solve,
and thus there is much more freedon.
This will be formalized in the last section.

\paragraph*{Plan of the paper.}
In section \ref{sec2},
we leave the binary tree setting
and reformulate the problem in terms of flips and triangulations,
essentially as in [2].
In section \ref{sec3},
we present the two far-apart triangulations
and study the sphere triangulation formed by their union.
In section \ref{sec4},
we define a weight function on triangles,
using the max-flow min-cut theorem and a partial order on triangles.
In section \ref{sec5},
we analyze the change of weights induced by flips,
and conclude the proof of the lower bound.
In section \ref{sec6},
we discuss connections between this proof,
linear programming,
and the original proof,
as well as other possible applications of the technique.

\section{Preliminaries : triangulations and tetrahedra} \label{sec2}

Rather than dealing with binary trees,
it is preferable to study the dual problem on triangulations,
which have more symmetry.

\subsection{Binary trees and triangulations} \label{sec2.1}

Consider a binary tree with $n$ internal nodes
and $(n+1)$ external leaves;
move the external leaves to infinity
by extending the edges to these leaves into half-lines;
add a distinguished half-line from the root to infinity.
The planar dual of this graph
is a triangulation of an $(n+2)$-gon
with one distinguished polygon edge.
This defines a bijection
between rooted binary trees with $n$ internal nodes
and triangulations of an $(n+2)$-gon
with a distinguished polygon edge.
A rotation on a binary tree
corresponds to a \emph{flip} in the triangulation,
obtained by replacing an edge
by the other diagonal of the quadrilateral
formed by the two triangles adjacent to this edge.
(See figure \ref{fig2}).
\figsize{350pt}{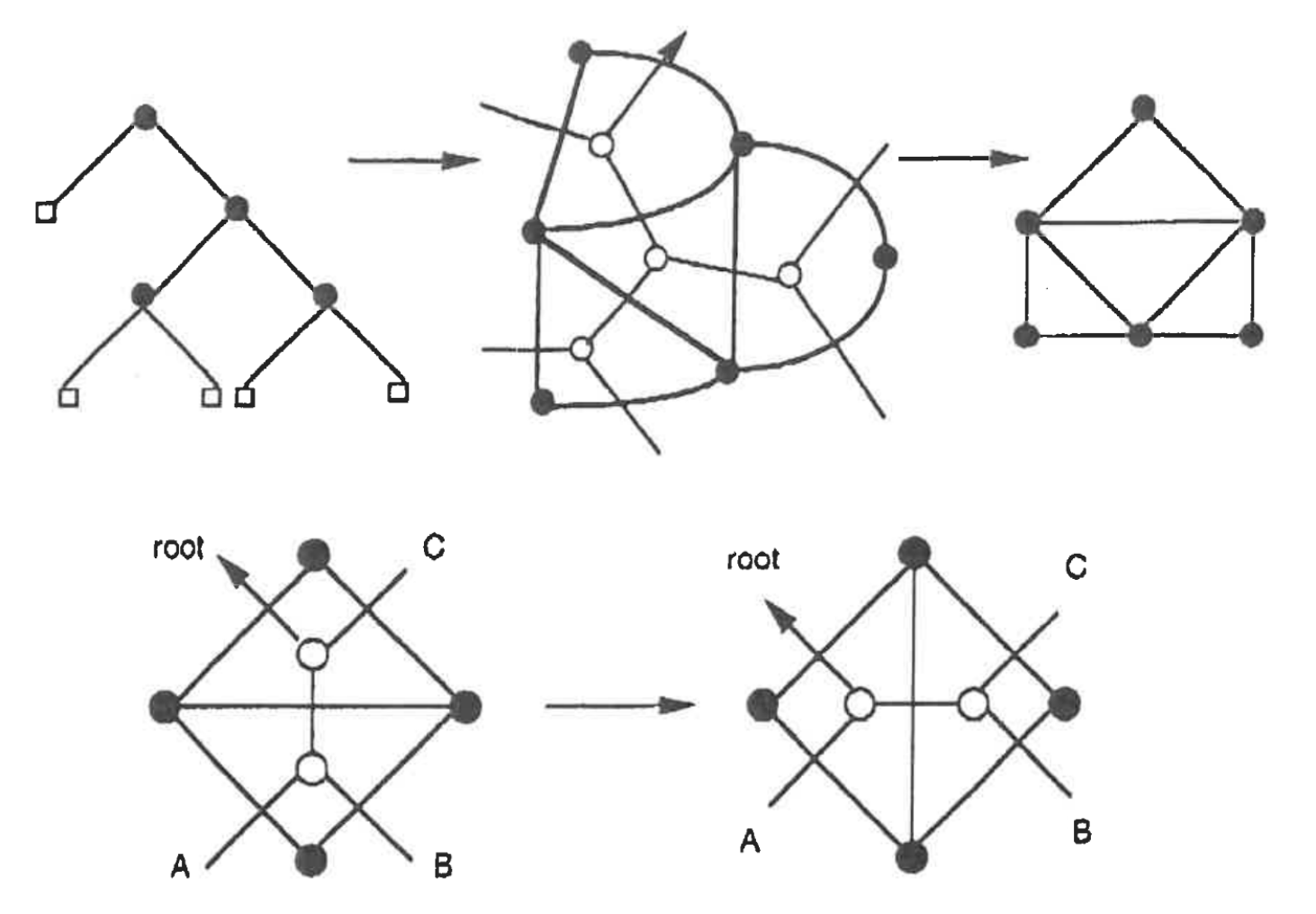}{fig2}{Binary trees and triangulations}
We define the \emph{flip distance}
between two triangulations $T_1$ and $T_2$
as the minimum number of flips
required to transform $T_1$ into $T_2$.
We will prove that the maximum flip distance
between triangulations of an $(n+2)$-gon is $2n-6$.

\subsection{Amortized analysis : defining a weight function} \label{sed2.2}

The proof idea is to define a weight function on triangles,
such that the weight of a triangulation
is the sum of the weights of its triangles;
such that any flip,
by destroying two triangles and creating two new triangles,
changes the total weight by at most $1$;
and such that there are two particular triangulations
whose weights differ by exactly $2n-6$.
The lower bound follows.
This is similar to the potential function arguments
used to prove amortized upper bounds
on the performance of dynamic algorithms.

Actually,
for the sake of symmetry,
we prefer to think in terms of oriented triangles.
An \emph{oriented triangle}
is a triple $\{i,j,k\}$
of polygon vertices (not necessarily adjacent),
together with one of two possible orientations,
$ijk=jki=kij$ or $ikj=kji=jik$,
such that changing the orientation multiplies by $-1$:
in other words,
$w(ijk)=-w(ikj)$ for all $i,j,k$.

Let $w(T)$
be the weight of a triangulation $T$
drawn inside the polygon
with all triangles oriented counterclockwise.
We draw the initial triangulation $T_i$
inside the polygon,
orienting all triangles counterclockwise.
We redraw the final triangulation $T_f$
so that all of its edges are outside the polygon,
orienting all triangles counterclockwise.
We obtain a triangulated planar graph $T$
with all triangles oriented counterclockwise
and with total weight $w(T)=w(T_i)-w(T_f)$.
A flip of triangles of $T_i$ or of $T_f$
can be seen as a flip in T
which preserves the counterclockwise orientation.
The central part of the proof
consists in showing
that any flip of any edge
changes the weight of the triangulation by at most $1$.

In particular,
transforming $T_i$ into $T_f$
by doing flips of edges inside the polygon
yields a graph with two copies of each triangle of $T_f$,
one inside and one outside the polygon,
with opposite orientations,
and thus the total weight is $0$.

Given a polygon with $(n+2)$ vertices,
the lower bound will be proved by exhibiting:
firstly, a weight function on oriented triangles
which satisfies the following \emph{tetrahedral constraints}:
for any quadruple $\{i,j,k,l\}$ of polygon vertices,
\begin{equation} \label{eq1}.
|w(ijk)+w(jlk)+w(jlk)+w(kli)+w(ilj)| \leq 1
,
\end{equation}
(which guarantees that any flip
changes the weight by at most $1$);
secondly,
a triangulated planar graph
which has a hamiltonian cycle
(i.e. which can be viewed as the union
of two polygon triangulations,
one on each side of the hamiltonian cycle)
and has weight $2n-6$
if all triangles are oriented counterclockwise.

Geometrically,
we can consider this planar triangulation
as a triangulation of the sphere
with all triangles oriented outwards.
Each flip corresponds to a tetrahedron.
In [2],
the idea is to
exhibit a sphere triangulation
which cannot be extended to a triangulation of the ball
with fewer than $2n-6$ tetrahedra.
Since the hyperbolic volume of any tetrahedron
is at most a constant $V_0$,
the proof focuses on constructing
a sphere triangulation
forming a polyhedron
of total volume greater than
$(2n-7)V_0$,
and almost all the work is devoted to
the difficult problem of calculating volumes in hyperbolic space.
From here on,
our proof differs from the original proof of [2].

\section{Choosing the two triangulations} \label{sec3}

\subsection{Two far apart triangulations} \label{sec3.1}

In this section,
we define two triangulations of an $(n+2)$-gon.
We will later prove that they are
at flip distance at least $2n-6$ apart.

The first one is the ``horizontal zig-zag'' triangulation,
and the second one is the zig-zag triangulation,
rotated by $\sqrt{n}$
(see figure \ref{fig3}).
\figsize{300pt}{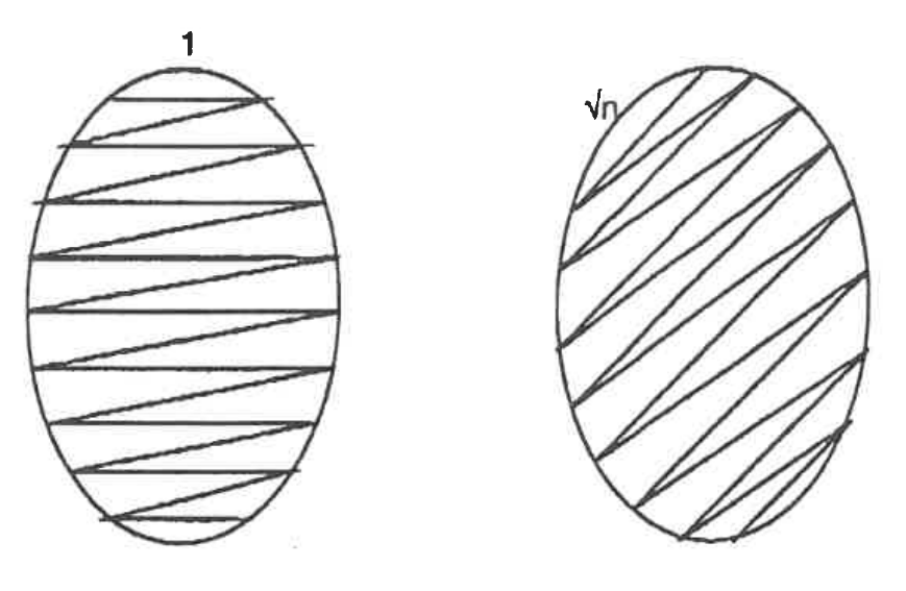}{fig3}{Two triangulations at distance $2n-6$ from each other}

In the planar triangulated graph $T$
obtained by drawing one triangulation inside
and the other one outside the polygon,
we note that most vertices have degree $6$:
they are adjacent to
two edges of the first triangulation,
two edges of the second triangulation,
and to two polygon edges.
The exceptions are the extremities of the zig-zags,
which make up for four vertices of degree $5$
and four vertices of degree $4$
in graph $T$.
In the neighborhood of a ``typical'' vertex,
the graph $T$ looks like a regular triangular lattice;
near the degree $4$ vertices,
the graph is as in figure \ref{fig4}.
\figsize{300pt}{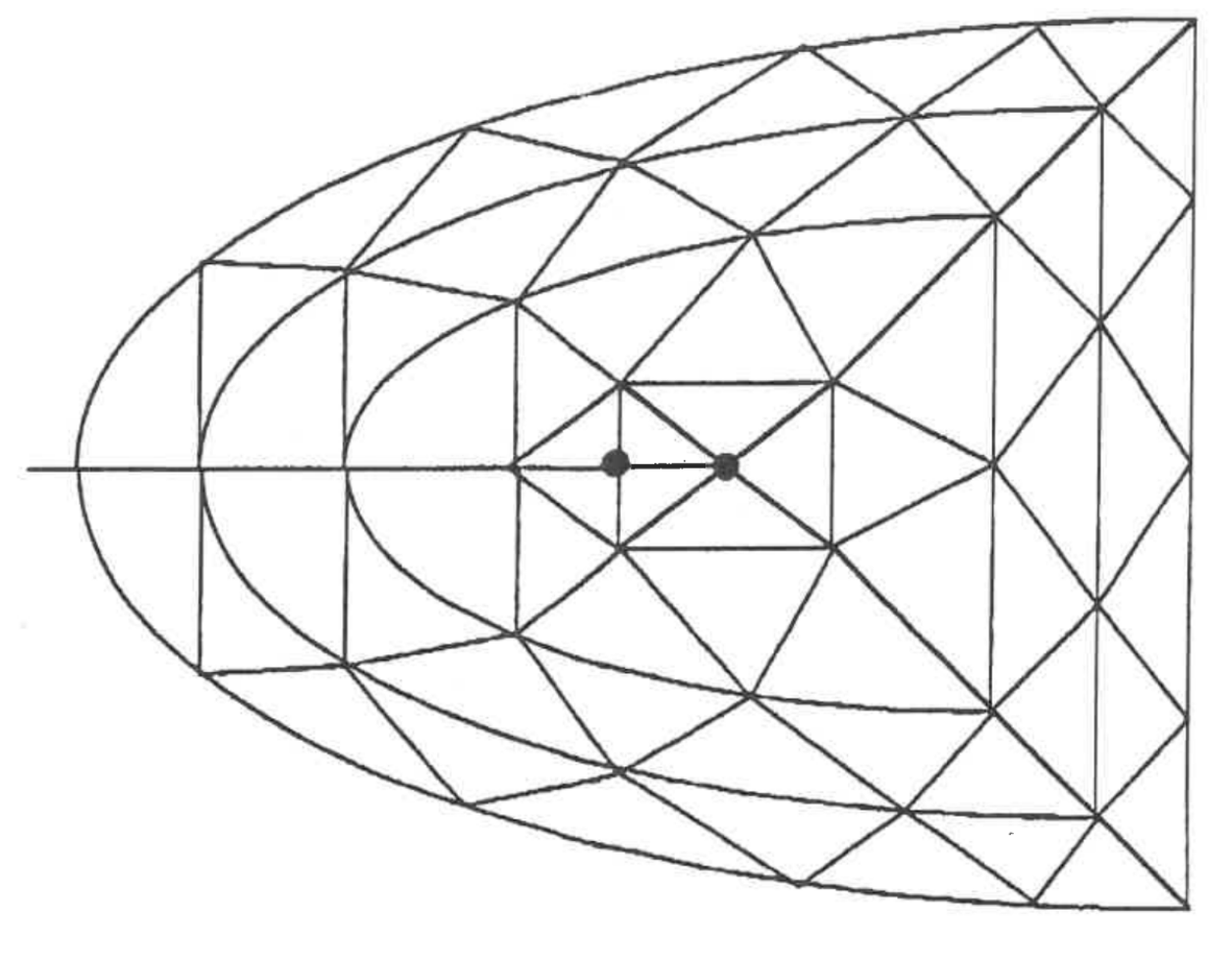}{fig4}
{The sphere triangulation in the neighborhood of a degree 4 vertex}
The $\sqrt{n}$ rotation angle was chosen
so that the degree $4$ vertices
are far from one another in $T$
as we will see later.
The weight function on oriented triangles
will be defined
%%%% hurray!
using graph $T$
and will require computing distances on this graph $T$.
Thus we need to understand distances on $T$.

\subsection{Visualizing distances on $T$} \label{sec3.2}
As a warmup,
let us start from the planar regular triangular lattice $L$,
and show how to use it to represent a triangulated graph $L'$
where every vertex has degree 6,
except for one degree 5 vertex and one adjacent degree 4 vertex,
in such a way that distances can easily be computed.

Pick a vertex $u$ from $L$;
its 6 edges extend to infinite half-lines
which partition the plane into 6 sectors.
Removing one sector and identifying the two half-lines bounding it
into one half-line $V$ decreases the degree of $u$ by 1.
Let $v$ be the neighbor of $u$ on $V$.
Removing the two sectors of $v$ on either side of $V$
and identifying their remaining boundaries decreases $v$'s degree by 2.
We obtain the following picture for $L'$,
where the arrows join vertices which are identified (figure \ref{fig5}).
\figsize{\linewidth}{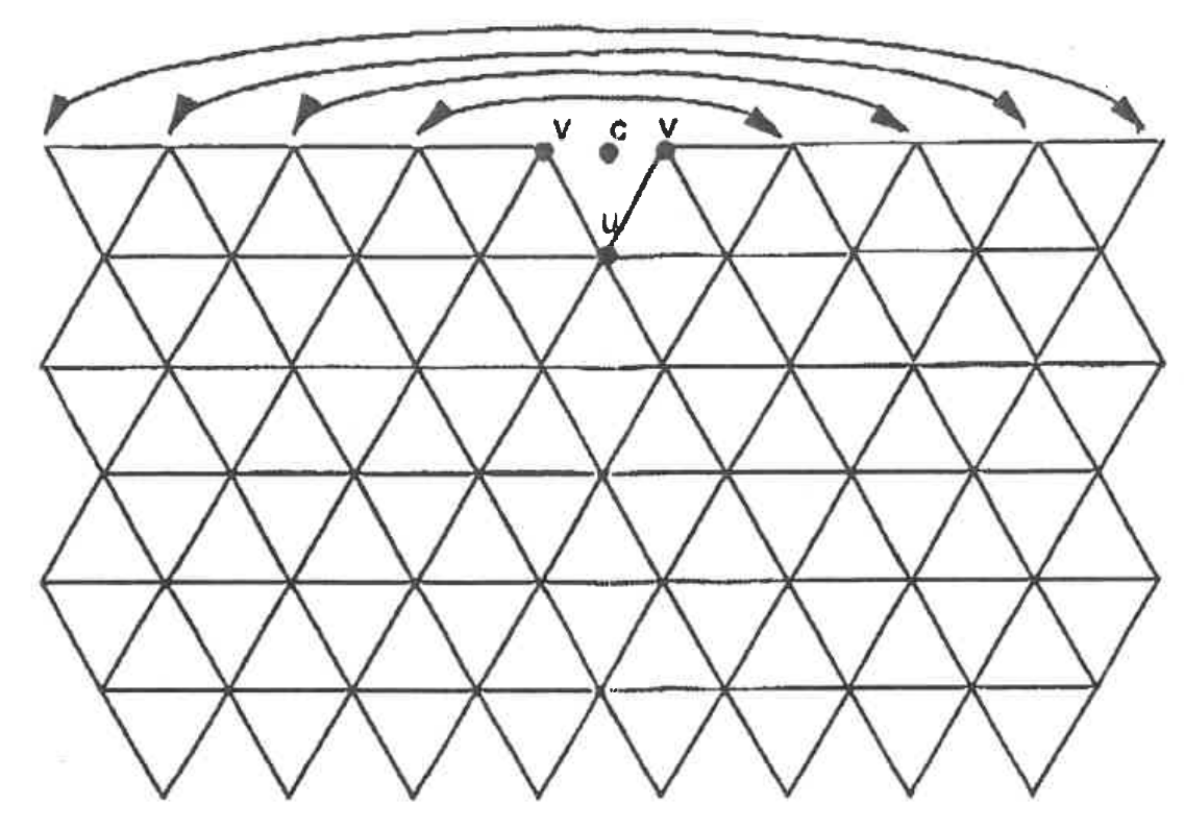}{fig5}{Representation of graph $L'$}
Let $c$ denote the center of the segment in $L$
joining the two points labeled $v$.
The triangular lattice $L$ is a double cover of $L'$
(except for the edge of $L$ going through $c$):
to compute the distance between two points $x$ and $y$ of $L'$,
we take $y'$, the image of $y$ by the symmetry of center $c$ in $L$,
and compute $d_{L'}(x,y)=\min(d_L(x,y), d_L(x,y'))$.
See figure \ref{fig6} for an example,
where the distance
is $d_{L'}(x,y)=\min(d_L(x,y),d_L(x,y'))=\min(6,4)=4$.
\figsize{\linewidth}{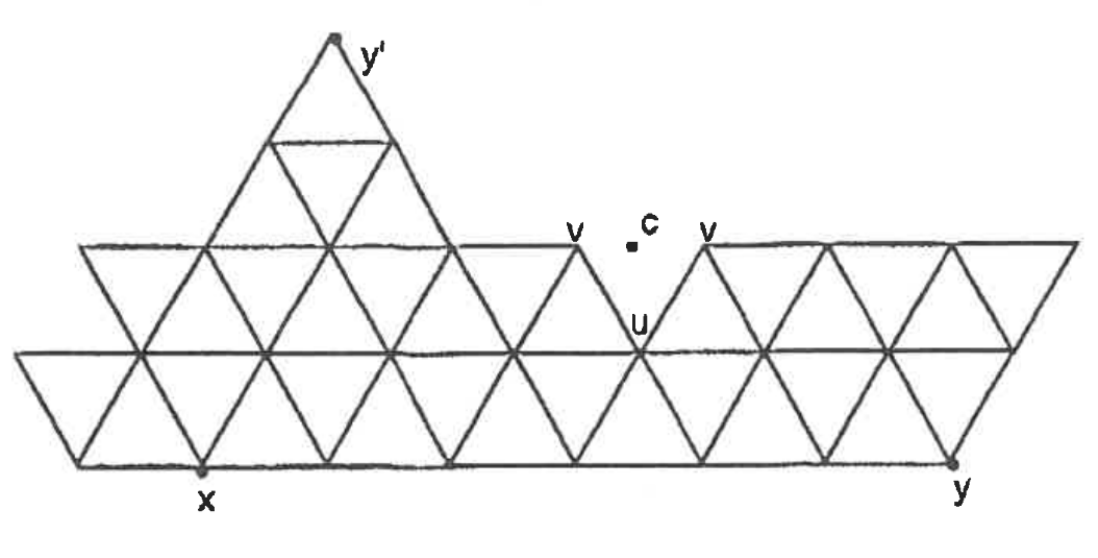}{fig6}
{Computing distances in $L'$ using the lattice  $L$}

We now turn to our problem of constructing
a convenient picture of $T$.
The same idea can be used,
using $L$ as an infinite cover.
We obtain the graph depicted in figure \ref{fig7}.
Let $c$ denote the center of any one of the black lozenges
(formed by two adjacent triangles of $L$).
Two vertices of $L$ which are images of each other
by the symmetry of center $c$
represent the same vertex of $T$.
A fundamental region consists of four big triangular regions,
which we label 0,1,2,3.
Graph $T$ can be recovered by taking such a fundamental region
(dotted in the figure)
and suitably identifying the sides
(identified vertices are joined by arrows in figure \ref{fig7}).
\figsize{\linewidth}{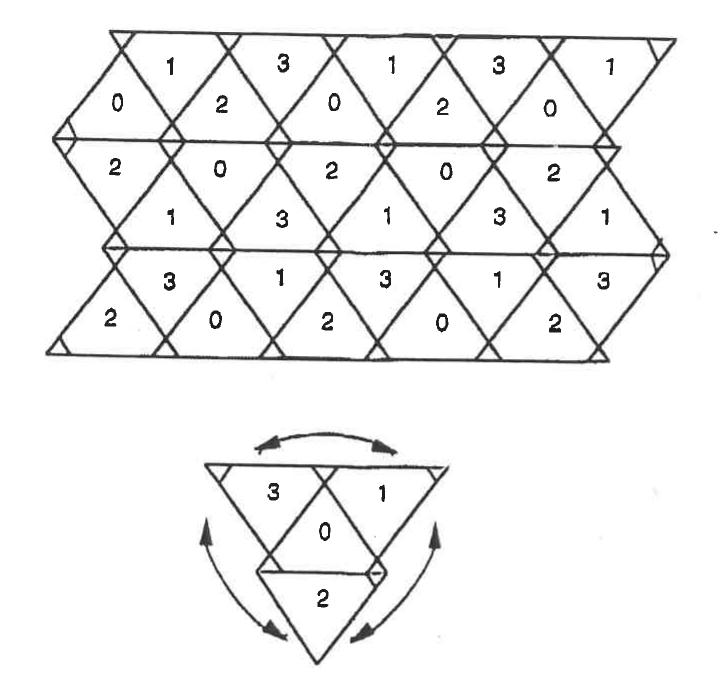}{fig7}{Computing distances in $T$ using $L$}

To compute the distance between vertices $x$ and $y$ in $T$,
we consider all the vertices $y_i$ of $L$ which represent $y$.
Then $d_T(x,y)=\min\{d_L(x,y_i)\}$.
In practice, for a given $x$ in $T$,
we take the set of all the representatives of $x$ in $L$,
and compute its Voronoi diagram (where distances are computed in $L$).
Then for any $y$, $d_T(x,y)=d_L(x_0,y_0)$,
where $x_0$ is an arbitrary (fixed) representative of $x$ and $y_0$
is the representative of $y$ in the Voronoi cell of $x_0$.

\section{Defining the weight function} \label{sec4}

The definition of the weight of a triangle $(ijk)$ will essentially depend on the position of its three vertices in $T$.

\subsection{Defining orientations of triangles} \label{sec4.1}
Before defining the actual value of the weight of triangles, we will first choose an orientation for each triangle $\{i,j,k\}$ such that the weight of the triangle will be non-negative for the chosen orientation. (Knowing the sign of $w(ijk)$, even if we do not know its exact value, will often be sufficient for checking the tetrahedral constraints \ref{eq1}).

In graph $T$,
consider non-crossing shortest paths
$p_{ij},p_{jk},p_{ki}$ from $i$ to $j$,
$j$ to $k$ and $k$ to $i$ respectively.
Their concatenation forms a cycle separating $T$
into two connected components.
The one with the least area
(i.e. with the fewest triangles)
is called the \textit{interior}.
We define the \textit{triangular region} $(ijk)$
to be the union of all the interiors of such cycles.
Informally, $(ijk)$ is the ``fattest" region
formed by geodesics between the vertices.
The boundary of this triangular region
usually defines a cyclic order on the vertices
(except if a vertex is in the interior of the triangular region,
or if the triangular region is flat,
or if the triangle is so big that
the triangular region contains the whole graph $T$).
We choose the orientation to be $ijk$
if the vertices occur in that order
in the counterclockwise direction around the boundary.
Also note
(we will use this later)
that inclusion of triangular regions
induces a natural partial order on triangles.
Unless otherwise specified,
from now on the ``weight of triangle $\{i,j,k\}$"
will refer to the triangle with this chosen orientation.

\subsection{Defining the weights of triangles} \label{sec4.2}
We partition triangles in classes,
depending on how many edges they have which are in the graph $T$.

Firstly, triangles of $T$:
the triangles which are not adjacent to
one of the degree $4$ or degree $5$ vertices have weight 1.
The triangles adjacent to a vertex of degree 4
have weight $3/4$.
For the triangles adjacent to a vertex of degree 5,
their weight, 3/4 or 1,
is defined as in figure \ref{fig8}.
\fig{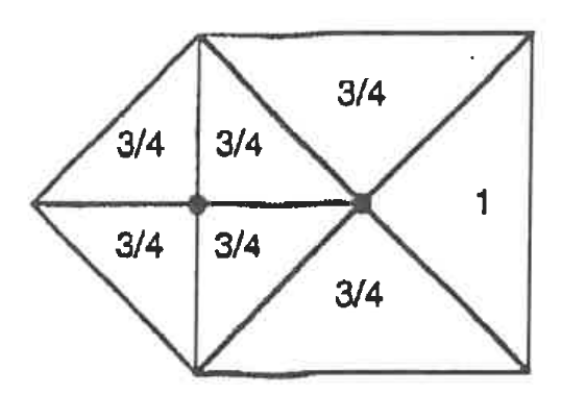}{fig8}{Defining weights near the special vertices}

Secondly, triangles having exactly two edges of $T$:
If the area of their triangular region is 0,
they have weight 0.
Otherwise the weight is defined
as in figure \ref{fig9}:
in most cases,
they are obtained by using two adjacent triangles
of weight $1$ in the flip, and are then given weight 1/2.
Other cases involve the degree 4 and degree 5 vertices,
and are listed in figure \ref{fig9}.
\figsize{\linewidth}{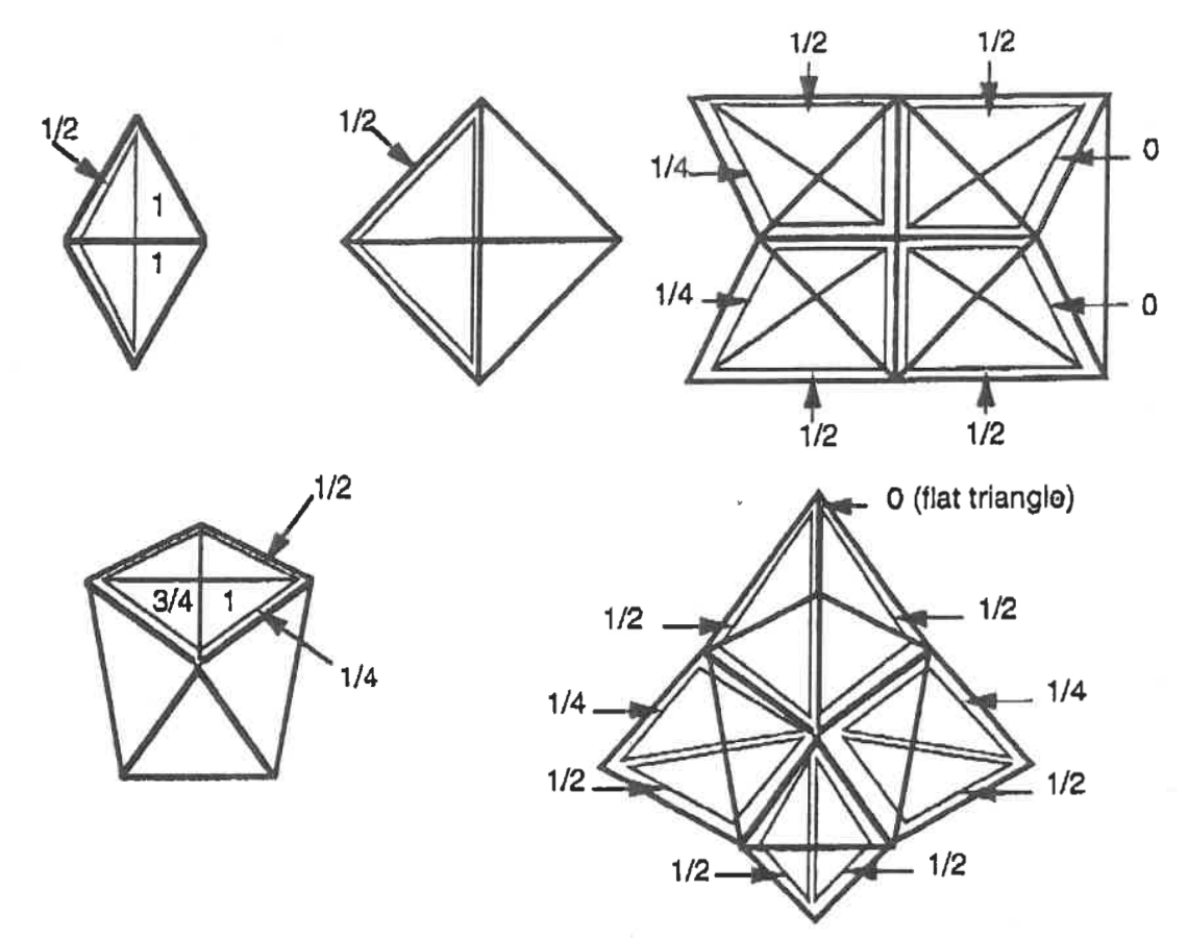}{fig9}{Weight of triangles created by one flip}

Thirdly, triangles having exactly one edge of $T$:
this is the crucial part of the construction.
Each such triangle is associated to its isolated vertex $s$
(the vertex which is not an extremity of the edge of $T$).
We will define the weight of all triangles
associated to $s$ simultaneously
by solving a flow problem (one flow problem for each $s$)
defined as follows.

We flip every edge opposite to $s$
in the six (or five or four in special cases) triangles of $T$
containing $s$.
(The degree of $s$ doubles in the process).
We consider the planar dual of the graph $T$ with these six flips, $T(s)$. Add a vertex $s_0$ to $T(s)$, the source for the flow problem,
linked to all the vertices dual to the triangles obtained by doing a flip.
Define the value of the source
as the sum of the weights of these triangles.
The sink for the flow problem is a new vertex $t_0$
linked to all the vertices dual to triangles of weight 3/4,
and each edge linked to the sink has capacity 1/4.

We now want to give an orientation to the edges of $T(s)$.
Let $e$ be an edge of $T(s)$
and $ab$ its dual edge of $T$.
If triangle $\{a,b,s\}$ is oriented,
then we direct $e$ \textit{outwards}
from the triangular region $(abs)$.

The next step is to define capacities.
We choose two constants $c$ large enough and $c'>10c$.
All oriented edges within distance $c$
of the source or of the sink are given capacity 1/2.
All oriented edges within distance $c'$ of the source
or of the sink are given capacity 1/4.
Forget the orientation of all the other edges of the graph,
and give them capacity $1/4$.
See figure \ref{fig10}.
\figsize{300pt}{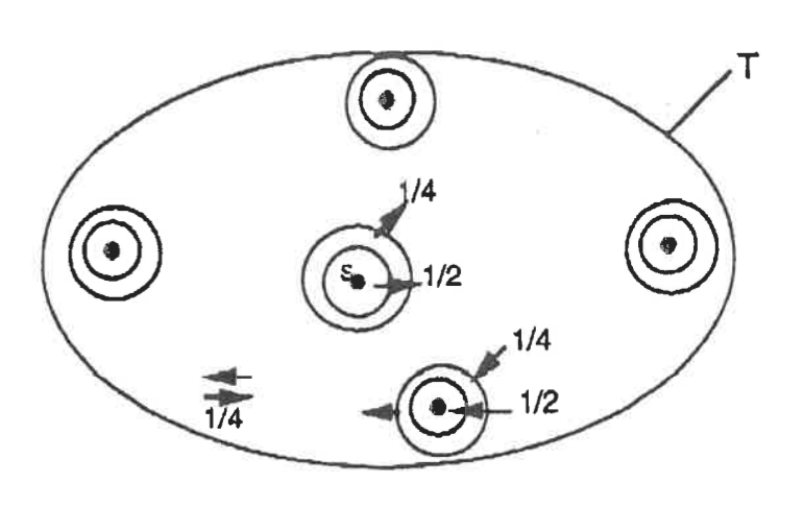}{fig10}{Defining capacities}

To show that there exists a flow
absorbing all the value coming out of the source,
we use the max-flow min-cut theorem
(in the straightforward direction),
along with the following lemma.
\begin{lemma} \label{lemma1}
Any cut in $T(s)\cup\{s_0\}\cup\{t_0\}$
separating the source $s_0$
from the sink $t_0$
has capacity at least equal to the value of the source.
\end{lemma}
This is proved by examining orientations, using the representation of $T$ described above (the proof will be in the full version).

Coming back to the problem of defining weights on triangles,
if $w$ is the flow through an edge $e$,
we give weight $w$ to the triangle $sab$
associated to $s$
and containing the edge $ab$ dual of $e$,
oriented so that the flow goes out of the triangle
and oriented counterclockwise.

An example is given in figure \ref{fig11}
for a ``typical" source (not in the neighborhood of the sink).
\figsize{\linewidth}{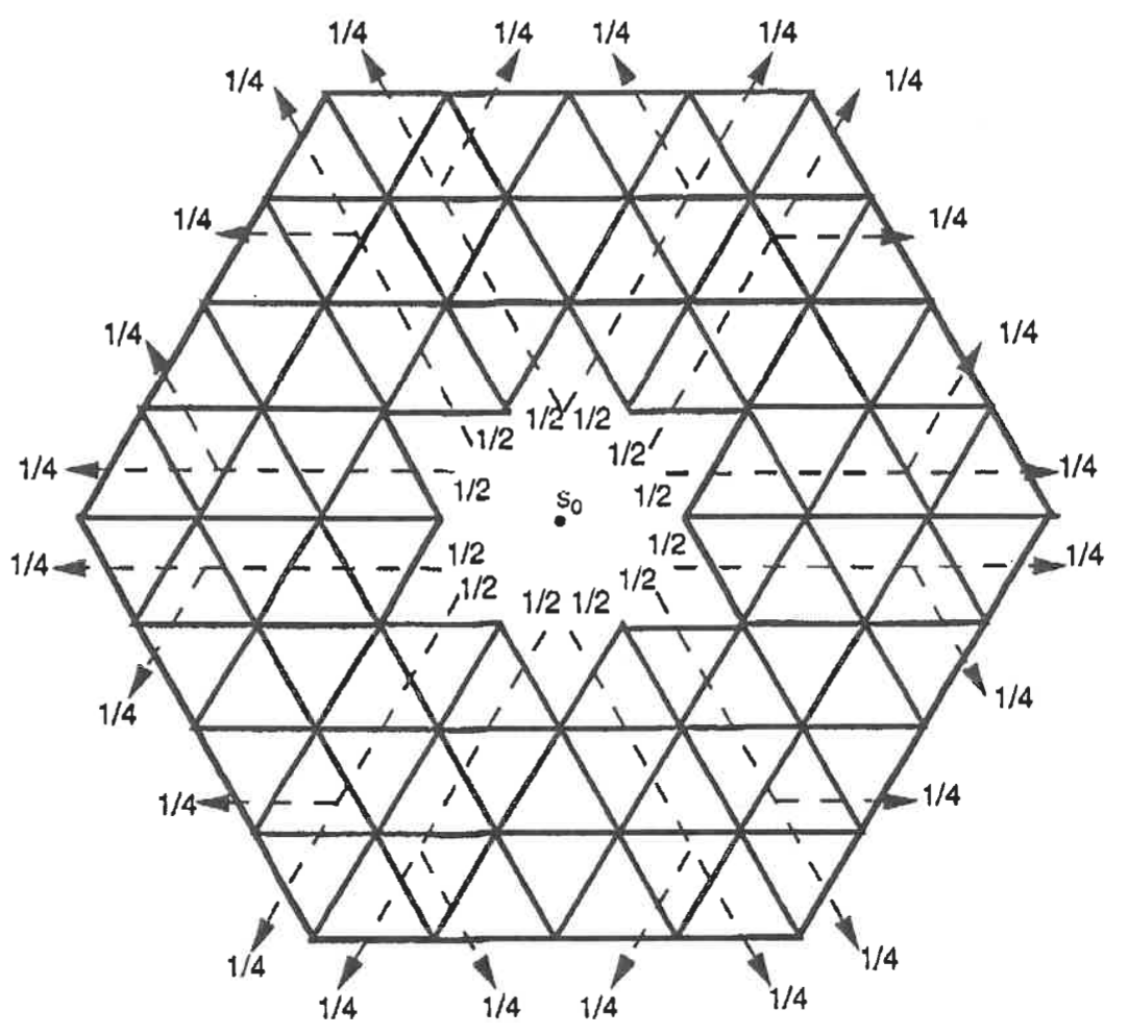}{fig11}{Defining weights using flows}

Fourthly, we need to define the weight of triangles with no edge of $T$:
We will define their weights inductively
in order of increasing area of their triangular region.
Area 0 triangles are given weight 0.
Let $(ijk)$ be a triangle of non-zero area.
We look at all vertices $l$ in the triangular region $(ijk)$
such that $(ijl),(jkl)$ and $(kil)$
all have area strictly smaller than the area of $(ijk)$,
and define a function $f(l)$:
$$f(l)=w(ijl)+w(jkl)+w(kil)-1.$$
We now define the weight of triangle $(ijk)$:
$$w(ijk)=\max\{0,\max_l f(l)\}.$$

This concludes the definition of triangle weights.

\section{Verifying the constraints} \label{sec5}

\subsection{The triangulation has weight $2n-6$} \label{sec5.1}
The graph $T$ has $2n$ triangles and $n+2$ vertices.
Every triangle of $T$ has weight 1
except in the neighborhood of vertices of degree less than 6:
there are 4 vertices of degree 4,
adjacent to 4 vertices of degree 5;
in the neighborhood of each of these 4 pairs,
there are 6 triangles of weight 3/4,
each accounting for a weight deficit of 1/4.
There is a total of 24 triangles of weight 3/4
and the total weight is thus $2n-24/4=2n-6.$

\subsection{Weights of large triangles} \label{sec5.2}
We want to check that each flip
decreases the weight by at most 1,
that is, that each quadruple $\{i,j,k,l\}$
(viewed as a tetrahedron with all triangles oriented outwards)
satisfies
\[
|w(ijl)+w(jkl)+w(kil)+w(kji)| \leq 1
.
\]
We will first show the following lemmas.
\begin{lemma} \label{lemma2}
If $\{i,j,k\}$ is not a triangle of $T$,
then $|w(ijk)|\leq 1/2$.\end{lemma}
{\bf Proof.}
If $\{i,j,k\}$ has two edges in $T$,
the proof is by inspection.
If it has one edge in $T$,
the result follows from the fact that
all capacities are at most 1/2.
If it has no edge in $T$, its weight, if non-zero,
is determined by some fourth vertex $l$.
By induction  $ijl, jlk, lki$
all have weight at most 1/2,
so $f(l)$ is at most $1/2+1/2+1/2-1\leq 1/2$.
Hence the lemma.

We will show that all large triangles have weight 0.

\begin{lemma} \label{lemma3}
If $d(i,j)>c+2$
and both vertices $i$ and $j$
are at distance greater than $c$ from the sink,
then the weight of $(ijk)$ is $|w(ijk)|\leq 1/4$.
\end{lemma}

{\bf Proof.} If $ijk$ has one edge in $T$, say $jk$, then this edge is far both from the source $i$ and from the sink, hence the dual has capacity 1/4 and $|w(ijk)|\leq 1/4$.

If $ijk$ has no edge in $T$,
the weight of $ijk$
is determined by doing a decomposition
into three triangles and iterating on each of these triangles
until they all have at least one edge in $T$
or have weight 0.
Let $ijl$ be the triangle of side $ij$
obtained in this decomposition.
If $ijl$ has weight 0,
working back up the decomposition,
we deduce $w(ijk)=0$.
If not,
then vertex $l$ is a neighbor either of $i$ or of $j$;
say that it is a neighbor of $i$.
Thus the edge dual to $il$,
being far from both the sink and the source $j$,
has capacity 1/4,
and $|w(ijl)|\leq 1/4$.
Working back up the decomposition,
we deduce $|w(ijk)|\leq 1/4$,
hence the lemma.

\begin{lemma} \label{lemma4}
Large triangles,
i.e.  such that
all three side lengths $d(i,j)$, $d(j,k)$ and $d(k,i)$
are greater than 10c, have weight 0.
\end{lemma}
The proof is similar to the proof of the previous lemma
and will be in the full version.

\subsection{Weights of tetrahedra} \label{sec5.3}
We will use these lemmas for analyzing the tetrahedral constraints \ref{eq1}.

Case 1: Tetrahedra with 2 faces in $T$: then all the faces are triangles with two or three edges in $T$. Proof by inspection.

Case 2: Tetrahedra with one face in $T$: the weight of the other three faces was defined using flows, precisely so that their contributions cancel. The total weight of the tetrahedron is thus at most 1.

Case 3: Tetrahedra with no face in $T$: this is more difficult. We look at the orientation of the four triangular regions defined by the faces of the tetrahedron.

Subcase 1: two positively oriented and two negatively oriented faces. A positive face must have nonnegative weight, except if it was defined using flows, unoriented and given capacity 1/4. If no face is in that special case, then there are two faces whose weight is between 0 and 1/2 and two faces whose weight is between -1/2 and 0: the total weight of the tetrahedron is thus between -1 and 1. If one of the faces, say $ijk$, is in the special case, then studying the possible positions of the fourth point $l$ of the tetrahedron and using $c'>3c$ proves that the total weight is at most 1,
by arguments similar to the proof of lemma \ref{lemma3}.
(Proof detailed in full version).

Subcase 2:
three positively oriented faces,
say $(ijl),(jkl),(kil)$
and one negatively oriented face $(kji)$.
If the area of $(ijk)$
is strictly greater than the areas of $(ijl),(jkl),(kil)$,
then by definition of the weight of triangle $(ijk)$,
the tetrahedron has weight at most one.
(The other cases will be in the full version).

Subcase 3:
four positively oriented faces.
Then at least two of these faces are huge
and satisfy the assumptions of lemma \ref{lemma4},
thus have weight 0.
The other two having weight at most 1/2,
the tetrahedron has weight at most 1.

This completes the proof of the lower bound.

\section{Comments} \label{sec6}
We note that the proof can be greatly simplified
for the more modest goal of proving the lower bound $2n-O(1)$.
Then, in section \ref{sec4}, when defining the weights of the triangles,
we can just give weight 0 to all the triangles of $T$
within some fixed neighborhood of the special vertices (degree 4 or 5).
The non-flat triangles with two edges of $T$
all have weight 0 in the neighborhood of the special
vertices and 1/2 otherwise.
The flow problem can be defined
by giving capacity 1/2 near the source and 1/4 everywhere else,
and the proof that there is no small cut becomes straightforward.
The proof of lemma \ref{lemma4} is greatly simplified,
and analyzing the tetrahedral constraints is also simpler.
The lower bound $2n-O(1)$ also holds for more general graphs,
namely any graph which looks like the triangular lattice
except for a constant number of local perturbations.

We conjecture that the same technique can be used
to characterize the pairs of polygon triangulations
which are at maximum distance apart.
In fact, in the sphere triangulation formed by their union,
it is known that there is no vertex of degree greater than 6.
Since the sum of the degrees of the $(n+2)$ vertices is $6(n+2)-12$,
there are only a finite number of cases to consider
for the degree sequence of the graph.
In some cases,
it is easy to see that the two triangulations
are not at maximum distance apart
(for example, if the sphere triangulation has a vertex of degree 3,
or two adjacent vertices of degree 4).
In the other cases,
we conjecture that the graph looks like the triangular lattice
(i.e. there is no short cycle separating the graph
into two large connected components:
the injectivity radius is not too small),
and that the same proof will work.

More generally,
the approach of using a potential function may be useful
for proving lower bounds on other problems involving structures
which evolve through local transformations.

This work stems from the last section of [2]
(in the journal version only),
which briefly sketched connections between the original proof
and linear programming.

More explicitly,
each flip of two triangles can be associated to
an \emph{oriented tetrahedron},
i.e. a quadruple
$\{i,j,k,l\}$,
vertices of the triangles involved,
together with a sign giving the direction of the transformation.
There are $N=2\binom{n+2}{4}$ oriented tetrahedra.
The distance between two triangulations $T_i$ and $T_f$ is the length
of the shortest path transforming $T_i$ in $T_f$.
To each such path,
we associate a vector $(x_1,\ldots,x_N)$,
where $x_j$
is the number of times that the flip
corresponding to tetrahedron $j$ occured on the path.
The length of the path is $x_1+\ldots+x_n$,
and all $x_j$'s are non-negative integers.
The constraint is that the path must lead from
$T_i$ to $T_f$.
Amazingly, this can be worded as a linear constraint,
using the linearity of the boundary operator
for simplices:
let $\del$ be the map
\[
\{\{i,j,k,l\},+\}
\mapsto
(ijl)+(jkl)+(kil)+(kji)
,
\]
where the image is a formal sum of oriented triangles
and we adopt the convention $(ijk)=-(kji)$.
Then the set of formal sums of triangles
is an $\binom{n}{3}$-dimensional vector space.
One can check that $\del$ can be extended
to a linear operator:
\[
\del \;:\;
(x_1,\ldots,x_N)
\mapsto
\sum_{1 \leq j \leq N}
x_j
\del (\text{tetrahedron $j$})
.
\]
The constraint that the path must go
from $T_i$ to $T_f$ then gives:
\[
\del ((x_1,...,x_N))
=T_f-T_i
,
\]
where $T_f$ (resp. $T_i$)
is the formal sum of triangles
of the triangulation $T_f$ (resp. $T_i)$,
with counterclockwise orientation.
If we solve the following linear program:
\[
m^*=\min (x_1+...+x_N)
\]
\[
s.t.\begin{cases}
\forall j, x_j\geq 0\\
\del(x_1,...,x_N)=T_f-T_i
,
\end{cases}
\]
we will have a lower bound $m^*$ on
the distance between $T_i$ and $T_f$.
The dual linear program is:
\[
M^*=\max{(T_f-T_i)}\cdot (w_1,...,w_{\binom{n}{3}})
\]
\[
s.t.\ \del^T w\leq (1,1,\ldots,1)
.
\]
Since $M^* \leq m^*$,
the value of $M^*$
is also a lower bound.

The constraints of the dual program
can be interpreted as follows:
for every tetrahedron, the sum of
the weights of the four triangles
bounding it is at most 1.
Thus this $w$ is precisely our weight function on triangles,
and our whole proof can be viewed
as constructing a feasible solution to the dual program.

In the hyperbolic geometry approach,
the authors choose an embedding
of the sphere triangulation in space,
with one vertex $v_0$ at infinity.
We can say that they define a weight function on triangles:
\[
w((i,j,k))=
\mathrm{Vol}(\{i,j,k,v_0\})/V_0
,
\]
where the volumes are computed in hyperbolic space.
If we consider the dual program,
in that case the constraints are trivially satisfied,
and the cost function is the volume
of the sphere triangulation
(up to a factor $1/V_0$).

It can be hoped that this linear programming
interpretation would help providing ``natural'' amortized
analyses for other problems.

\section*{References}

[1] Daniel D. Sleator and R. E. Tarjan.
\textit{Self-adjusting binary search trees}, J.ACM 32, 1985, 652-686.

\noindent
[2] D. D. Sleater, R. E. Tarjan and W. P. Thurston.
\textit{Rotation distance, triangulations, and hyperbolic geometry},
Journal of the AMS, 1, 3, 1988, 647-681. (Preliminary version in STOC 1986).

\end{document}